\documentclass[a4paper]{spie}

\usepackage[]{graphicx,color}
\usepackage{textcomp}

\title{Modeling the electromagnetic properties of the SCUBA-2 detectors} 

\author{Michael D. Audley\supit{a}, Giampaolo Pisano\supit{b}, Wayne S. Holland\supit{a}, William D. Duncan\supit{c}, William Parkes\supit{d}, and Peter A. R. Ade\supit{b}
\skiplinehalf
\supit{a}UK Astronomy Technology Centre, Royal Observatory, Edinburgh, EH9 3HJ, UK; \\
\supit{b}Cardiff University, Cardiff, CF24 3YB, UK; \\
\supit{c}National Institute of Standards and Technology, 325 Broadway, Boulder, CO 80305, USA; \\
\supit{d}The Scottish Microelectronics Centre, University of Edinburgh, Edinburgh, EH9 3JF, UK
}

\authorinfo{Further author information: (Send correspondence to M.D.A.)\\M.D.A.:
 E-mail: mda@roe.ac.uk, Website: http://www.roe.ac.uk/atc/projects/scubatwo/}

\begin{document} 
  \maketitle 

{\bf Copyright 2004 Society of Photo-Optical Instrumentation Engineers.\\}
This paper will be published in SPIE conference proceedings volume 5498, 
``Millimeter and Submillimeter Detectors for Astronomy II.''  and is made available as 
an electronic preprint with permission of SPIE.  One print or electronic 
copy may be made for personal use only.  Systematic or multiple reproduction, 
distribution to multiple locations via electronic or other means, duplication 
of any material in this paper for a fee or for commercial purposes, or 
modification of the content of the paper are prohibited.

\begin{abstract}
SCUBA-2 is the next-generation replacement for SCUBA (Sub-millimetre 
Common User Bolometer Array) on the James Clerk Maxwell Telescope.  Operating at
 450 and $850\rm\ \mu m$, SCUBA-2 fills the focal plane of the telescope with 
fully-sampled, monolithic bolometer arrays.  Each SCUBA-2 pixel uses a 
quarter-wave slab of silicon with an implanted resistive layer and backshort as an absorber
 and a superconducting transition edge sensor as a thermometer.
In order to verify and optimize the pixel design, we have investigated the 
electromagnetic behaviour of the detectors, using both a simple
transmission-line model and Ansoft HFSS\texttrademark, a finite-element 
electromagnetic 
simulator.  We used the transmission line model to fit transmission measurements
 of doped wafers and determined the correct implant dose for the absorbing 
layer.
The more detailed HFSS modelling yielded some unexpected results which led us 
to modify the pixel design.  We also verified that the detectors suffered 
little loss of sensitivity for off-axis angles up to about $30^\circ$.
\end{abstract}




\newcommand {\Section}[1]{Section~\ref{#1}}

\newcommand {\Figure}[1]{Figure~\ref{#1}}

\newcommand {\Table}[1]{Table~\ref{#1}} 

\newcommand {\page}[1]{page~\pageref{#1}}

\newcommand {\Equation}[1]{Equation~\ref{#1}}

\newcommand {\arcdeg}{^\circ}
\def\plotfiddle#1#2#3#4#5#6#7{\centering \leavevmode
\vbox to#2{\rule{0pt}{#2}}
    \includegraphics{#1}}


\def\captionfigure#1[#2]#3{
 \def\captionlabel{#1}
 \def\captionlistentry{#2}
 \def\captionheading{#3}
 \begin{figure}}

\def\endcaptionfigure{
 \spacing{1}
 \caption [\captionlistentry]{\captionheading}
 \label {\captionlabel}
 \end{figure}}

\def\sqr#1#2{{\vcenter{\vbox{\hrule height.#2pt
        \hbox{\vrule width.#2pt height#1pt \kern#1pt
          \vrule width.#2pt}
        \hrule height.#2pt}}}\relax}
\def\square{\mathchoice\sqr45\sqr45\sqr{3.1}4\sqr{2.5}4}

\section{INTRODUCTION}

\begin{figure}[hb]
\hskip1in
\input{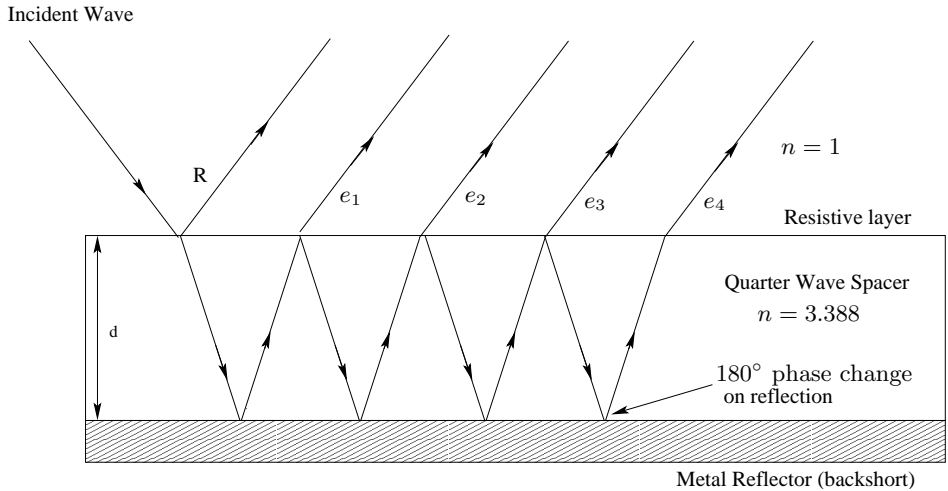}
   \caption
   { \label{salisburyscreen}    
Operation of the SCUBA-2 pixels.  The detectors consist of a metal reflector and a resistive layer spaced a quarter wavelength apart by a slab of silicon.  This arrangement is known as a Salisbury screen.  The emergent waves ($e\,_i$) interfere destructively with the reflected wave (R) and constructively with the incident wave.  This maximizes the energy dissipated in the resistive layer at the top of the quarter-wave slab and minimizes the amount reflected.} 
\end{figure}

In SCUBA-2\cite{SPIE04} two detector arrays will observe simultaneously at wavelengths ($\lambda$) of 850 and $450\rm\ \mu m$.  Each array consists of four $40\times32$ sub-arrays of square pixels.  Each pixel comprises a silicon slab which has a metal layer, or {\it backshort}, behind it.
For the interface between free space and the front of the slab, the reflection coefficient is
\begin{equation}
R=\left({1-n\over 1+n}\right)^2
\end{equation}
where $n$ is the refractive index of the slab.  We assume $n=3.388$ for cold silicon, giving $R=30\%$.  However, we can reduce the reflectivity (and thus maximize the detector efficiency) if the 
thickness of the slab is chosen so that the radiation reflected by the backshort at the 
back of the slab interferes destructively with the radiation reflected by the top of the slab.  In both cases, the radiation undergoes a $180\arcdeg$ phase change on reflection, so the slab thickness is chosen to be an odd multiple of $\lambda/4$ in the slab material.
This arrangement is known as a Salisbury screen 
 and is used in aviation for reducing radar reflections.  The operation of the detector is shown in Figure~\ref{salisburyscreen}.  At sub-millimetre wavelengths and low temperatures, silicon has negligible dissipation.
Thus, the slab has an implanted resistive layer at the top to dissipate the energy of the radiation, heating up the pixel.  
To maximize the detector efficiency, we need to choose the implantation dose so that the surface resistance is $R_s\approx377\ \Omega/\square\,$, matching the wave impedance of free space ($Z_0=120\,\pi\rm\ \Omega$).   The temperature rise due to the absorption of radiation is 
detected by a transition-edge sensor (TES) on the back side of the quarter-wave slab.  It is this TES that acts as the reflective backshort.  

The pixel geometry is repeated across the detector wafer to form one of SCUBA-2's $40\times32$ sub-arrays.
Each sub-array is bump-bonded to a wafer that carries the SQUID multiplexing (MUX) circuitry.  The indium bump bonds provide electrical and thermal connections, as well as mechanical strength.

The size of an individual quarter-wave slab ($1055\rm\ \mu m$, allowing for the wall thickness and the gap between the walls and slab) is on the order of the wavelength of the incident radiation (850 and $450\rm\ \mu m$).
This means that to design the pixels for maximum efficiency we need a model of the electromagnetic behavior of the detector.  We need answers to the following questions.  How should we design the pixels 
to maximize the detector efficiency, and how much variation in the detector parameters can we tolerate?  
How does the efficiency of the detector depend on the wavelength, polarization, and incident angle?
To answer these questions, we carried out electromagnetic modeling.  We approached the problem in two ways.  First, we modeled the detectors as a simple transmission line. This model is equivalent to an infinite dielectric slab absorber and does not take into account any effects due to the finite size of the pixels, the supporting grid, or the periodicity of the array.  Then, we used a commercial software package, Ansoft HFSS\texttrademark\ (High-Frequency Structure Simulator; see {\tt http://www.ansoft.com}), to model the behaviour of the detectors more realistically.  HFSS is a three-dimensional finite-element method electromagnetic simulator.

\section{TRANSMISSION LINE MODEL}
\label{transmissionlinesection}
\begin{figure}
\hskip2in
\input{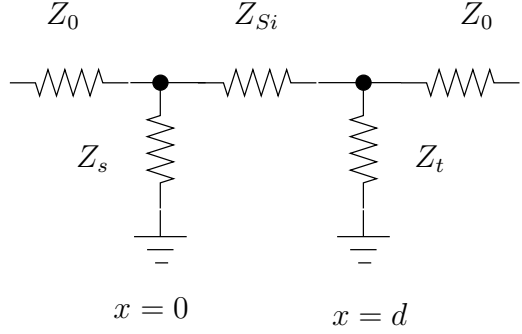}
   \caption
   { \label{transmissionline}    
SCUBA-2 quarter-wave slab represented by a simple transmission line model. The slab has a thickness $d$.  The surface impedances of its front and rear faces are $Z_s$ and $Z_t$.  The dielectric of refractive index $n$ is represented by an impedance $Z_{Si} =Z_0/n$, where $Z_0$ is the impedance of free space.  
For the SCUBA-2 pixels we have $R_s = 377\rm\ \Omega / \square \,$ 
and $Z_t\to0$.}
\end{figure}

For a simple model we can represent the vacuum, silicon, metal sandwich as a transmission line with corresponding impedances (see \Figure{transmissionline}).  Because we are assuming that the pixels extend laterally to infinity, we neglect any effects due to the finite size of the pixels or the periodicity of the array.  These effects will be addressed in \Section{software}.
Using this simple transmission line model, the voltage transmission coefficient is 

\begin{equation}
\label{t}
t = {4\,e^{i\,4\pi\,n\,d\,\sigma}\,n\,{z_s}\,{z_t}\over
     \left( 1 - \left(n-1 \right) \,{z_s} \right) \,
          \left( \left( n-1 \right) \,{z_t} - 1\right) 
        + e^{i\,4\pi\,n\,d\,\sigma}\,\left( 1 + {z_s} + n\,{z_s} \right) \,
        \left( 1 + {z_t} + n\,{z_t} \right)}
\end{equation}
and the voltage reflection coefficient is
\begin{equation}
\label{r}
r = {\left( {z_s} + n\,{z_s} -1\right) \,
        \left( \left( n-1 \right) \,{z_t} -1\right)  - 
       e^{i\,4\pi\,n\,d\,\sigma}\,\left( \left( n-1 \right) \,{z_s} +1\right) \,
        \left( 1 + {z_t} + n\,{z_t} \right)\over
          \left( 1 - \left( n-1 \right) \,{z_s} \right) \,
          \left( \left( n-1 \right) \,{z_t} -1\right)
        + e^{i\,4\pi\,n\,d\,\sigma}\,\left( 1 + {z_s} + n\,{z_s} \right) \,
        \left( 1 + {z_t} + n\,{z_t} \right)}
\end{equation}
where $z_s=Z_s/Z_0$ and  $z_t=Z_t/Z_0$.  The corresponding power transmission, reflection, and absorption coefficients are $T=|t|^2$, $R=|r|^2$, and $A=1 - T - R$, respectively.  In the general case, these expressions are complicated.

\begin{figure}
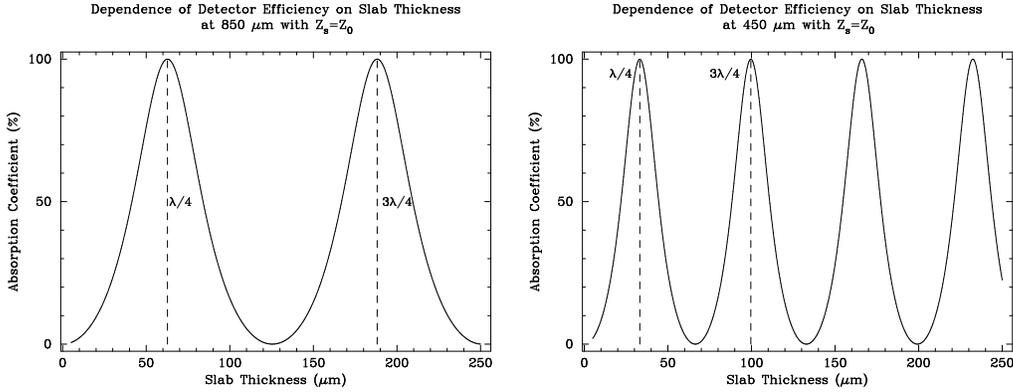

{\centering \leavevmode \vbox to144.465pt{\rule{0pt}{144.465pt}
\includegraphics{randa_850um_bw.qdp.epsi}
\includegraphics{randa_450um_bw.qdp.epsi}
}}
   \caption
   { \label{resistiveabsandrefl}
Dependence of quarter-wave slab absorption coefficient on slab thickness predicted from the simple transmission line model for wavelengths of 850 (left) and  $450\rm\ \mu m$ (right).  The quarter-wave slab has a resistive layer on the front with $R_s=Z_0$.} 
\end{figure}

\begin{figure}
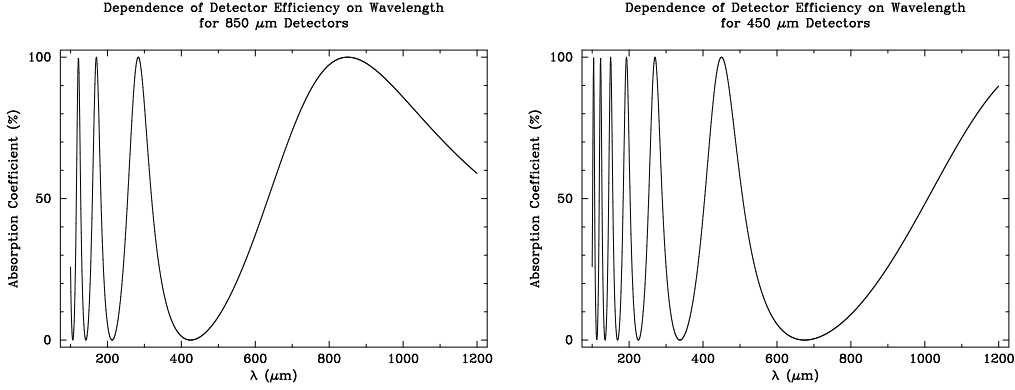

{\centering \leavevmode \vbox to143.207pt{\rule{0pt}{143.207pt}
\includegraphics{efficiency_vs_lambda_850um.qdp.epsi}
\includegraphics{efficiency_vs_lambda_450um.qdp.epsi}
}}
   \caption
   { \label{efficiencyvswavelength}    
Wavelength dependence of absorption coefficient of quarter-wave slab predicted from the simple transmission line model.  Left: Slab thickness $d=62.7\rm\ \mu m$ ($850\rm\ \mu m$ detector).  Right: Slab thickness $d=99.6\rm\ \mu m$ ($450\rm\ \mu m$ detector). The quarter-wave slab has a resistive layer on the front with $R_s=Z_0$.} 
\end{figure}

To model the behaviour of the SCUBA-2 detectors, we let $z_t\to0$ in \Equation{t}.  This means that we are assuming that the back of the slab is covered by a perfect conductor (the backshort).  Figures~\ref{resistiveabsandrefl} and \ref{efficiencyvswavelength} show how the predicted absorption coefficients depend on the slab thickness and the wavelength of the incident radiation.  We are assuming that the implanted absorbing layer is purely resistive.  Note that the ${3\over4}\lambda$ slab of the $450\rm\ \mu m$ detectors has greater wavelength selectivity than the ${1\over4}\lambda$ slab of the $850\rm\ \mu m$ detectors.


\subsection{Angular Dependence}
In the current design, the SCUBA-2 optics are not telecentric.  Pixels at the edges of the arrays will have rays incident on them at angles up to about $18\arcdeg$.  Thus, we must ensure that the efficiency of the detectors does not fall off too rapidly with incident angle.  Otherwise, pixels near the edges of the arrays will be insensitive.  We used the simple transmission line model to investigate the angular dependence of the detector efficiency.    We generalized the model to oblique incidence by making the substitutions shown in \Table{substitutions}\cite{Hadl47,Carli81}.  For normal incidence, the two linear polarizations are equivalent.  However, for oblique incidence, we must treat the s-polarization (E-field parallel to the plane of incidence) and the p-polarization (E-field normal to the plane of incidence) separately.

\begin{table*}[h]
 \centering
 \begin{minipage}{140mm}
  \caption{Substitutions to generalize transmission line model to case of oblique incidence.  $\Re(Z_{Si})$ is the real part of $Z_{Si}$.}
  \begin{tabular}{@{}lccc@{}}
  \hline
Quantity&Normal\hfill&\multispan2\hfill Oblique Incidence\hfill\\
&incidence\hfill&&\\
&&s-polarization&p-polarization\\
\hline
Free-space impedance&$Z_0$&$Z_0\cos\theta$&$Z_0/\cos\theta$\\
Slab impedance&$Z_{Si}$&$Z_{Si}\cos(\arcsin({\Re(Z_{Si})\sin\theta\over Z_0}))$&${Z_{Si}/\cos(\arcsin({\Re(Z_{Si})\sin\theta\over Z_0}))}$\\
Slab thickness&$d$&$d\cos(\arcsin({\Re(Z_{Si})\sin\theta\cos\theta\over Z_0}))$&$d\cos(\arcsin({\Re(Z_{Si})\sin\theta\cos\theta\over Z_0}))$\\
\label{substitutions}
\end{tabular}
\end{minipage}
\end{table*}

Using the substitutions in \Table{substitutions} with the transmission line model we calculated the angular dependence of the efficiencies of the SCUBA-2 detectors for s- and p-polarized radiation.  The results are shown in \Figure{simpleangulardependence}.  We found that the ${3\over4}\lambda$ slab of the $450\rm\ \mu m$ detectors has greater polarization-dependence than the ${1\over4}\lambda$ slab of the $850\rm\ \mu m$ detectors.  

\begin{figure}
{\centering \leavevmode \vbox to146.287pt{\rule{0pt}{146.287pt}
\includegraphics{850um_abs.qdp.epsi}
\includegraphics{450um_abs.qdp.epsi}
}}
   \caption
   { \label{simpleangulardependence}    
Angular dependence of absorption efficiency of SCUBA-2 $850\rm\ \mu m$ (left) and $450\ \mu m$ (right) detectors for s- and p-polarized  radiation, predicted from the simple transmission line model.} 
\end{figure}

\subsection{Effect of Filter Bandpass}
The radiation incident on the SCUBA-2 detectors will not be monochromatic.  To investigate the effect of the finite bandpass of the SCUBA-2 filters, we smeared the results of the transmission line model over a double-Gaussian filter function with a frequency bandpass of 10\%.  The effect of smearing over the filter bandpass is slight in the case of the $850\rm\ \mu m$ detectors.  However, for the $450\rm\ \mu m$ detectors the on-axis efficiency is degraded to 93.35\%\ of its monochromatic value.  The results are shown in \Figure{smearedangulardependence}.  The difference between the 850 and $450\rm\ \mu m$ detectors is not unexpected.  As was shown in \Figure{efficiencyvswavelength}, the $450\rm\ \mu m$ detectors have greater wavelength selectivity.

\begin{figure}
{\centering \leavevmode \vbox to144.676pt{\rule{0pt}{144.676pt}
\includegraphics{smeared850um_abs.qdp.epsi}
\includegraphics{smeared450um_abs.qdp.epsi}
}}
   \caption
   { \label{smearedangulardependence}    
Angular dependence of absorption efficiency of SCUBA-2 $850\rm\ \mu m$ (left) and $450\rm\ \mu m$ (right) detectors for s- and p-polarized  radiation, predicted from the simple transmission line model assuming a 10\%\ frequency bandpass (modeled as a double Gaussian).  The on-axis efficiency is degraded to 99.14\%\ of its monochromatic value for the $850\rm\ \mu m$ detectors and 93.35\%\ for the $450\rm\ \mu m$ detectors.} 
\end{figure}

\subsection{Implant Dose}

\begin{figure}
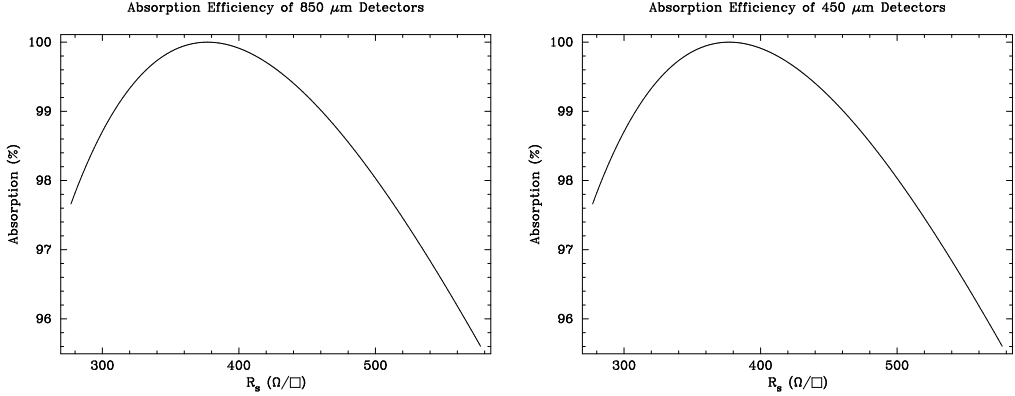

{\centering \leavevmode \vbox to147.092pt{\rule{0pt}{147.092pt}
\includegraphics{efficiency_vs_Rs_850um.qdp.epsi}
\includegraphics{efficiency_vs_Rs_450um.qdp.epsi}
}}
   \caption
   { \label{rsdependence}    
Dependence of absorption efficiency of SCUBA-2 850 (left) and  $450\rm\ \mu m$ (right) detectors on surface resistance $R\,_s$ predicted from the simple transmission line model.} 
\end{figure}

Our first application of the transmission line model to real data was to determine the correct implant dose for the implanted layer that absorbs the incident radiation.  
Wafers were implanted on the front side with different doses.  The implantation process is described elsewhere in these proceedings\cite{Park04}.  The submillimetre transmission of the wafers was measured at 1.5~K as a function of wavenumber ($\sigma$) using a Fourier transform spectrometer at Cardiff University.  
To model the transmission of these implanted wafers, we let $z_t\to\infty$ in \Equation{t}.  
We found that the model gave a satisfactory fit only if we added another component to the model --- an overall normalization factor --- and allowed this to vary freely.  The fits are consistent with the implanted layer being purely resistive rather than metallic.
Assuming we have a purely resistive implanted layer, as long as the correct thickness is chosen for the quarter-wave spacer the absorption efficiency does not depend strongly on the surface resistance $R\,_s$ (see \Figure{rsdependence}).

\section{FINITE-ELEMENT MODELING WITH ANSOFT HFSS}
\label{software}
\subsection{Description of the Software}
The simulations described here were carried out using HFSS version 8.5.  HFSS includes a three-dimensional drawing program that can be used to construct a geometrical model.  In other parts of the software, the user 
specifies the material properties, boundary conditions, and electromagnetic excitation of the model.
HFSS divides the volume of the model into a mesh of tetrahedra and solves for the electric field on the mesh in the frequency domain.  
The problem is converted into a set of linear equations, represented by a matrix. 
A field solution is obtained by inverting the matrix for the initial mesh. The mesh is then refined, based on the local field strength. The field solution from the refined mesh 
is compared with the previous solution.  This process of refinement and solving continues until the user-specified 
convergence criteria are met.  The convergence criteria depend on the type of excitation.  For  the models described here the model is excited with an incident planewave and convergence is evaluated from changes in the total scattered energy, calculated from the E-field.  

HFSS includes tools for 
visualizing the field solutions and calculating physical quantities from them.  These may be run from scripts to calculate quantities such as the reflection and absorption coefficients automatically.
We used the Ansoft 
Optimetrics\texttrademark\ package to run HFSS repeatedly on models while varying geometric or material 
parameters.

\subsection{Description of the Models}
\subsubsection{Method}
We simulated an infinite array of pixels by modeling a single square pixel and applying periodic boundary 
conditions to opposite walls.  The models are excited by an incident plane wave.  HFSS solves for the E-field 
in the model.  From the E-field we can calculate the physical quantities that interest us.  These are the power absorbed in the resistive layer, the total incident and reflected power, and the power flowing in or out of the unit cell through the side walls.  
The power absorbed in the implanted resistive 
layer is calculated from the field on the surface: 
\begin{equation}
P_{abs}={1\over2 R_s}\int_S ({\bf E}\times{\bf \hat n})^2 dS
\end{equation}
where $R_s$ is the surface resistance of the implanted layer and $\bf \hat n$ is the unit normal to the surface.  We have assumed $R_s = 377\rm\ \Omega/\square\,$
for all of the models.  The other powers are found by integrating the Poynting vector over the appropriate surfaces.

\begin{figure}[ht]
\plotfiddle{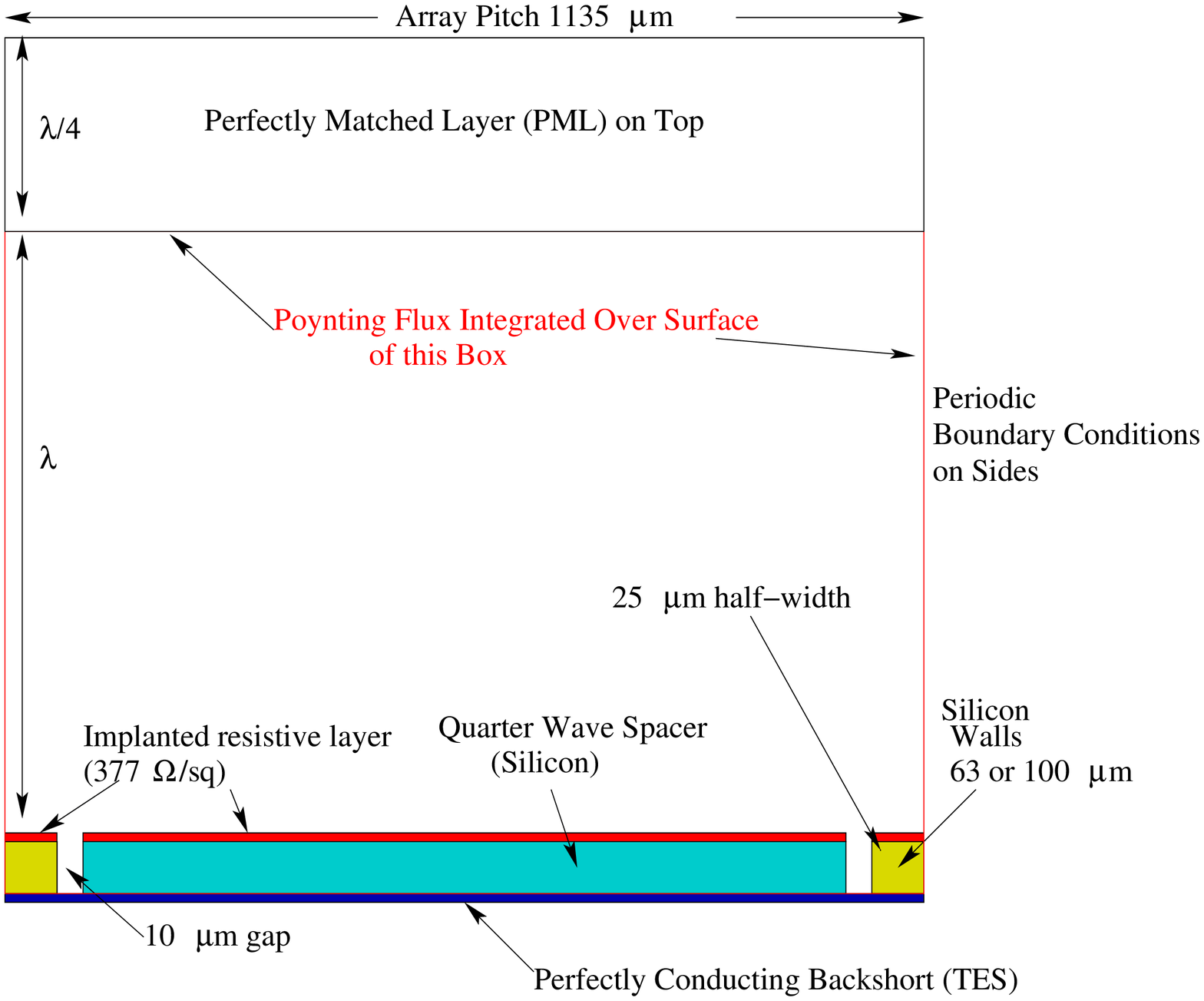}{300.929pt}{0}{58.2}{58.2}{-183.060pt}{-0.000pt}
   \caption
   { \label{current}    
Pixel model used in HFSS simulations.  
}
\end{figure}

Our basic pixel model consists of an approximation to a unit cell of the array (see \Figure{current}).   Vertical boundaries run through the 
centres of the silicon walls.  The walls are $50\rm\ \mu m$ thick, so that the unit cell contains the half-width of $25\rm\ \mu m$.  Periodic boundary conditions are applied to opposite side walls.  In the original pixel design there was a supporting grid $381\rm\ \mu m$ high on top of the walls.  This was later removed from the design (see \Section{grid}).  The results presented here are for models without this grid, unless otherwise stated.  The volume of the model extends $1{1\over4}$ wavelengths above the top of the walls (or grid, if present).  Inside 
the walls is the quarter-wave silicon slab with a resistive layer on its top surface.  Under the slab is a perfectly conducting layer, to simulate the superconducting TES.  

\subsubsection{Selection of boundary conditions}

For the top of the model where the waves come in, we have a choice of radiation boundary or perfectly 
matched layer (PML).  Both of these are designed to emulate free space by absorbing incident radiation.  The radiation boundary will absorb normally-incident radiation but it has a finite reflectivity for large angles of incidence.  The PML is a lossy, anisotropic dielectric that will 
also absorb radiation incident at large angles.  Because we wanted to illuminate the pixels with radiation incident at oblique angles, we chose a PML for the top of the model.  

On the side walls, we apply periodic boundary conditions for consistency with the plane wave excitation.  The  periodic boundary conditions on the pixel walls 
result in an asymmetric matrix, which increases the problem size.  However, neither radiation boundaries or PMLs are suitable for the side walls and yield results inconsistent with energy conservation.

\subsubsection{Error Analysis}

Because HFSS provides no direct measure of the errors in the results, we needed to find some way to check the quality of solutions.
The results can be checked for consistency by accounting for all of the incident power flowing into the unit cell.  The integrated flux 
flowing through all the walls of the model plus the power absorbed in the resistive layer of the
quarter-wave slab should add up to the incident power.  Any difference should be small for a high-quality solution.  In the plots below, the error bars on the absorption and reflection coefficients are derived from this discrepancy in the total power.
While they are not useful for deriving statistical uncertainties, they are shown so that the relative accuracy of the solutions can be compared.  
Another way to evaluate the solution is to calculate the incident power flowing through the upper boundary 
and compare it with the Poynting flux integrated over the surface.  The results should be slightly different 
due to the finite size of the mesh elements.  In HFSS, for plane wave excitation, the default setting is to 
apply a free-space plane wave with a peak electric field of $1\rm\  V\ m^{-1}$.  From this we can calculate the incident 
flux as
\begin{equation}
P_{inc}=\int_A{\bf S.dA}={1\over2}{\bf E}\times{\bf H}\,A\cos\theta={1\over2}\sqrt{\epsilon_0\over\mu_0}E^2\,A\cos\theta
\label{calculatedflux}
\end{equation}
where $A$ is the area of the top surface of the model, and $\theta$ is the angle of incidence.   The factor of 1/2 arises from the fact that the HFSS 
field calculator uses peak phasors to represent the fields.  We found that the incident flux calculated from \Equation{calculatedflux} sometimes differed from the integrated Poynting flux by up to 10\%.
Thus, in the results given below, the incident flux is derived from \Equation{calculatedflux}.

\begin{figure}
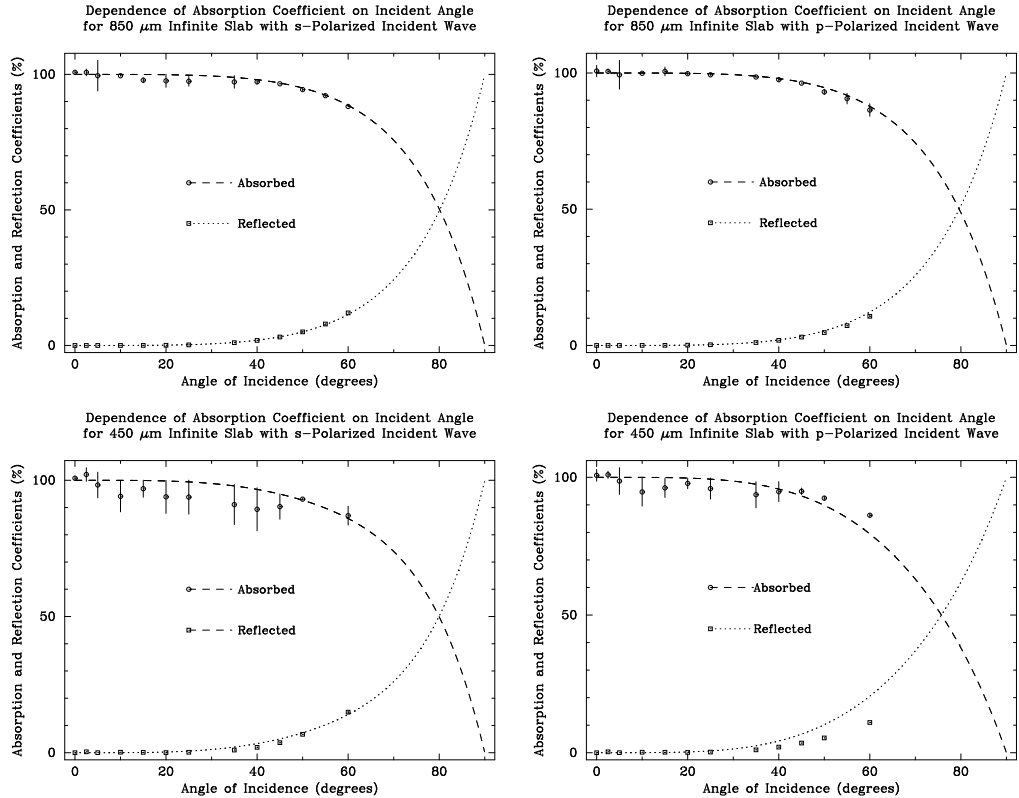

{\centering \leavevmode \vbox to144.676pt{\rule{0pt}{144.676pt}
\includegraphics{salisbury_screen_850_s.qdp.epsi}
\includegraphics{salisbury_screen_850_p.qdp.epsi}
}}
\vskip.3cm
{\centering \leavevmode \vbox to144.676pt{\rule{0pt}{144.676pt}
\includegraphics{salisbury_screen_450_s.qdp.epsi}
\includegraphics{salisbury_screen_450_p.qdp.epsi}
}}
   \caption
   { \label{comparison}    
Comparison of results from HFSS (points) and the simple transmission line model (lines) for a Salisbury screen.} 
\end{figure}

In order to check the accuracy of the HFSS solutions we simulated the 850 and $450\rm\ \mu m$ detectors as infinite slabs and compared the results with those from the transmission line model.  \Figure{comparison} shows good agreement for both, although the $450\rm\ \mu m$ efficiency is underestimated and has greater uncertainty.  This is because the shorter wavelength requires a finer mesh for the same precision.  The ultimate limit on the precision of the field solution is memory.  This limits the total number of tetrahedra in the mesh.  The version of HFSS we used is a 32-bit program.  This means that it will fail if the mesh grows to more than 4 GB.  In order to obtain the best precision possible, we pushed the mesh refinement until this limit was reached when solving our models.

\section{RESULTS OF THE HFSS MODELING}
\label{results}
\subsection{Detection Efficiency}
\begin{figure}
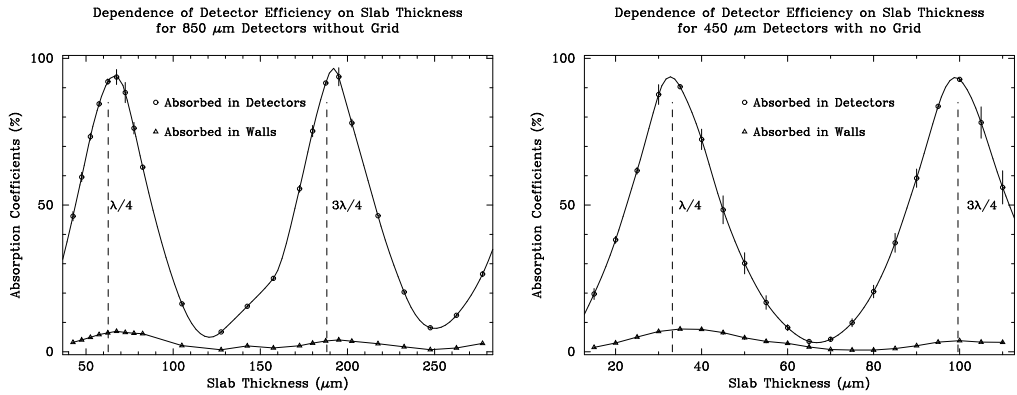

{\centering \leavevmode \vbox to144.676pt{\rule{0pt}{144.676pt}
\includegraphics{thickness850_grid0_bw.qdp.epsi}
\includegraphics{thickness450_grid0_bw.qdp.epsi}
}}
   \caption
   { \label{thickness}    
Predicted variation of detector efficiency with quarter-wave slab thickness for $850\rm\ \mu m$ (left) and $450\rm\ \mu m$ (right) detectors from HFSS simulations.  The vertical dashed lines correspond to the expected thicknesses for maximum efficiency.} 
\end{figure}

We used the Ansoft Optimetrics package to investigate how the absorption coefficient varied with the thickness of the quarter-wave slab.  The results are shown in \Figure{thickness}.  For the $850\rm\ \mu m$ detectors the absorption coefficient is greatest with a slab thickness of $\sim65\rm\ \mu m$.  From the assumed value of the refractive index ($n=3.388$) we would expect the maximum to occur at a slab thickness of $62.7\rm\ \mu m$ which corresponds to $\lambda/4$ in silicon.  For the $450\rm\ \mu m$ detectors the maximum occurs close to the expected value of $100\rm\ \mu m$ ($3\lambda/4$).  In each case the peak monochromatic detection efficiency was 94\%.

\subsection{Angular dependence}
\begin{figure}
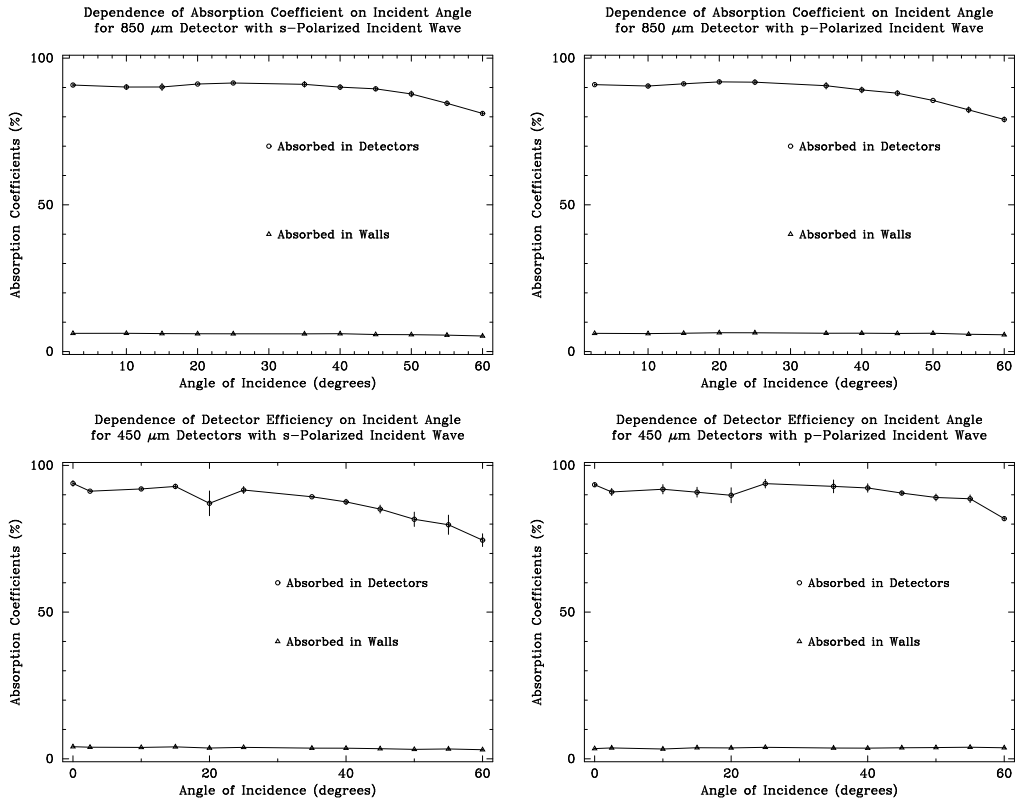

{\centering \leavevmode \vbox to144.676pt{\rule{0pt}{144.676pt}
\includegraphics{scuba2_oblique_l_850_50umwalls_s.qdp.epsi}
\includegraphics{scuba2_oblique_l_850_50umwalls_p.qdp.epsi}
}}
\vskip.3cm
{\centering \leavevmode \vbox to144.676pt{\rule{0pt}{144.676pt}
\includegraphics{scuba2_oblique_l_450_s.qdp.epsi}
\includegraphics{scuba2_oblique_l_450_p.qdp.epsi}
}}
   \caption
   { \label{angular}    
Predicted variation of $850\rm\ \mu m$ (top) and $450\rm\ \mu m$ (bottom) detector efficiency with angle of incidence from HFSS simulations.  Left: s-polarization.  Right: p-polarization. }
\end{figure}

We investigated the off-axis response of both the 850 and $450\rm\ \mu m$ detectors with the Ansoft Optimetrics package.
Neither detector suffers much loss in sensitivity for incident angles up to $30\arcdeg$ (see \Figure{angular}).  However, this sensitivity to oblique radiation does make the detectors susceptible to stray light.

\subsection{Effect of Supporting Grid}
\label{grid}
\begin{figure}[ht]
\hskip2in
\scalebox{.8}{\input{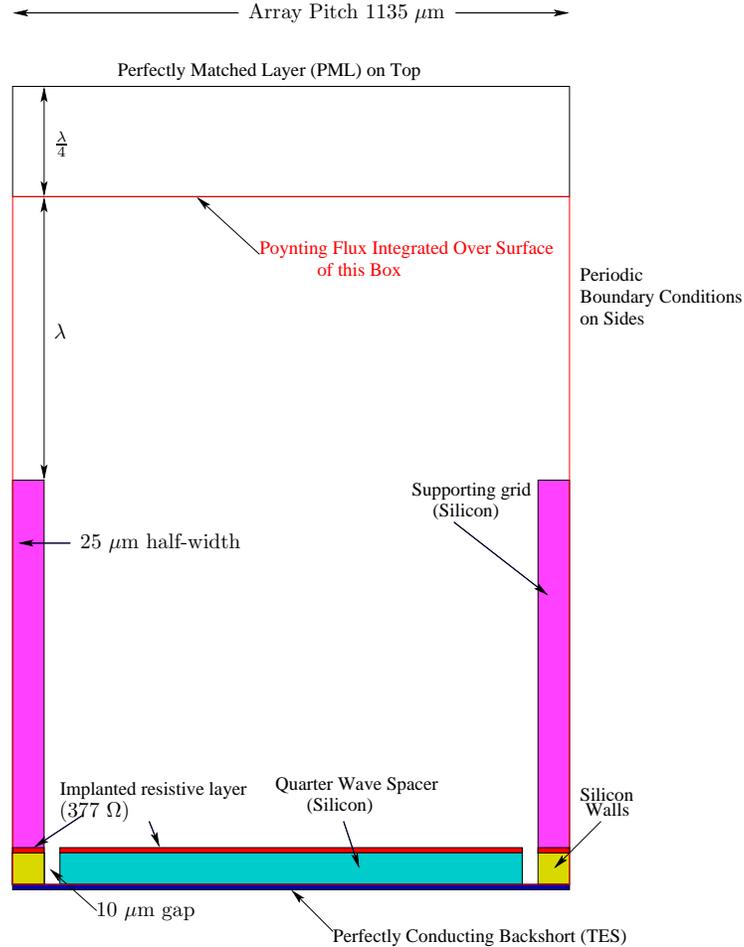}}
   \caption
   { \label{withgrid}    
Pixel model corresponding to original pixel design.  Note that the silicon support grid has been removed from the design and is omitted from simulations described here, unless otherwise stated.}
\end{figure}

\begin{figure}
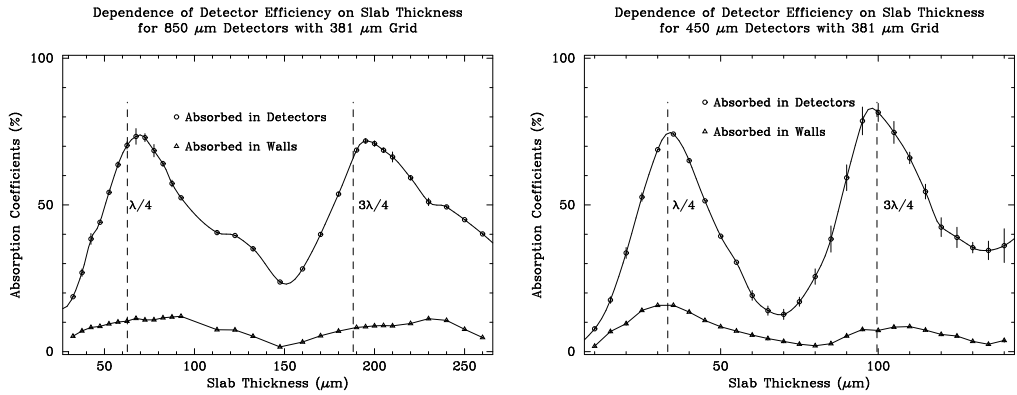

{\centering \leavevmode \vbox to144.676pt{\rule{0pt}{144.676pt}
\includegraphics{thickness850_grid381_bw.qdp.epsi}
\includegraphics{thickness450_grid381_bw.qdp.epsi}
}}
   \caption
   { \label{thicknesswithgrid}    
Predicted variation of detector efficiency with quarter-wave slab thickness for $850\rm\ \mu m$ (left) and $450\rm\ \mu m$ (right) detectors from HFSS simulations that include a $381\rm\ \mu m$ mechanical support grid.  The vertical dashed lines correspond to the expected thicknesses for maximum efficiency.} 
\end{figure}

The original design for the SCUBA-2 pixels had a silicon grid supporting the detector wafers.  A model that includes this grid is shown in \Figure{withgrid}.  This grid was essentially an extension of the side walls by $381\rm\ \mu m$, a standard wafer thickness.  The purpose of the grid was to provide mechanical rigidity to the detector array.  Since silicon is essentially transparent at these wavelengths, its effect on the detector performance was thought to be insignificant.  However, the HFSS modeling revealed that the grid acted as a dielectric waveguide, both reducing the detector efficiency and changing the optimum slab thickness (see \Figure{thicknesswithgrid}).

\begin{figure}
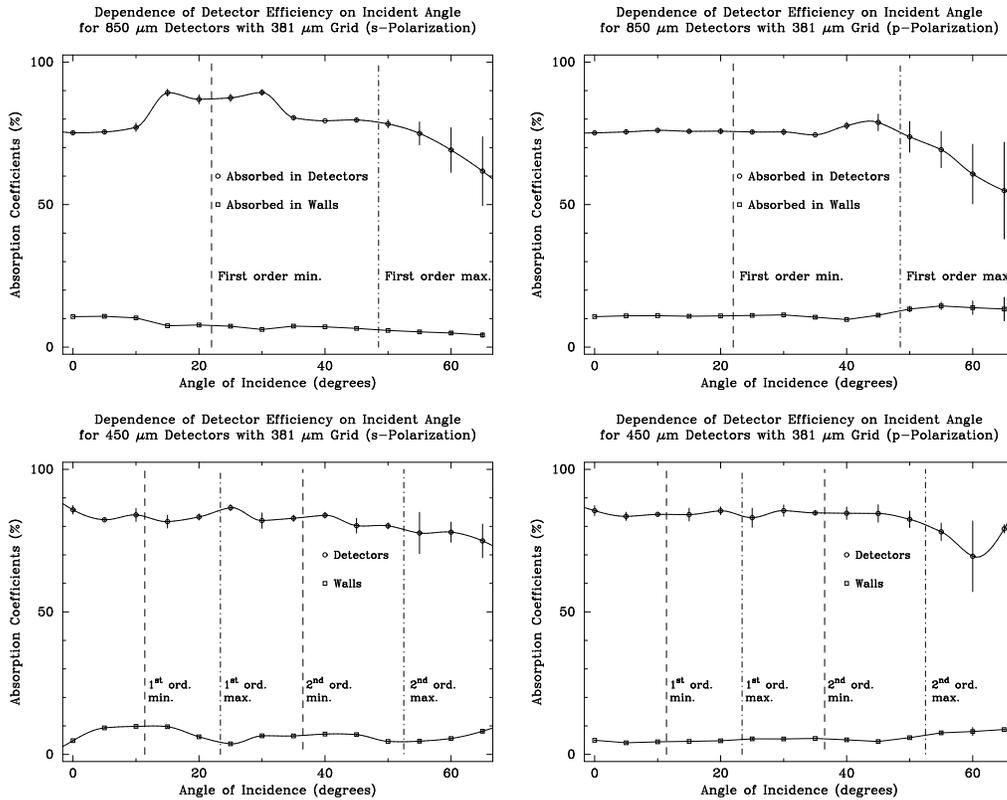

{\centering \leavevmode \vbox to144.676pt{\rule{0pt}{144.676pt}
\includegraphics{angle_850_s_bw.qdp.epsi}
\includegraphics{angle_850_p_bw.qdp.epsi}
}}
\vskip.3cm
{\centering \leavevmode \vbox to144.676pt{\rule{0pt}{144.676pt}
\includegraphics{angle_450_s_bw.qdp.epsi}
\includegraphics{angle_450_p_bw.qdp.epsi}
}}
   \caption
   { \label{angularwalls}    
Predicted variation of $850\rm\ \mu m$ (top) and $450\rm\ \mu m$ detector efficiency with angle of incidence from HFSS simulations that include a $381\rm\ \mu m$ mechanical support grid.  The vertical dashed lines represent the diffraction minima and maxima for the array pitch ($1135\rm\ \mu m$).  Left: s-polarization.  Right: p-polarization.}
\end{figure}

The mechanical support grid also affects the angular dependence of the detector efficiency (see \Figure{angularwalls}).  The grid structure behaves like a two-dimensional diffraction grating when the wavelength is smaller than the grid spacing. At both wavelengths (450 and $850\rm\ \mu m$) the array allows diffraction orders to propagate. This means that the non-absorbed reflected field is a superposition of the zeroth order (specular) component plus all the diffraction orders that can be generated. These depend on the frequency, angle of incidence and the grid spacing (i.e. the array pitch).  HFSS models this behaviour in a region that is close to the array, i.e. in its near field.  This means that we must be careful when comparing the simulations with experimental data.  When we measure the reflected flux we can be in a far-field configuration where only the zeroth order is detected. In this case the HFSS results have to be processed to calculate the far-field in the periodic boundary case. The important point is that the method we have described to calculate the absorption remains correct.

Because of its effect on the detector efficiency and because  changes in the processing had made it redundant anyway we removed the mechanical support grid from the design.

\subsection{Flatness of quarter-wave slab}

We used the Ansoft Optimetrics package to investigate the effect of the top of the quarter-wave slab not 
being flat.  To the top of the slab we added a wedge, and adjusted the thickness of the slab 
and that of the wedge, keeping the mean thickness of the combined structure constant.  This was done so 
that we could see the effects of having the top of the slab not parallel to the backshort, independent of 
changes in the thickness of the slab.

The top of the quarter-wave slab can make an angle of up to $1.6\arcdeg$ with the backshort before reducing the absorption 
coefficient for the p-polarization by 5\%.  This corresponds to a deviation in the slab thickness of up to $\pm15\rm\ \mu m$ across a pixel.

\subsection{Effect of oxide layer}

The fabrication process requires a layer of silicon oxide on the slab and walls as an etch stop.  We ran simulations with a $1\rm\ \mu m$ thick layer of silicon dioxide on top of the walls and the slab.  It had no significant effect on our results.

\section{Conclusions}
\label{conclusions}
\begin{itemize}
\item The optimum quarter-wave slab thicknesses to attain at least 95\%\ of the maximum efficiency are $66\pm6\rm\ \mu m$ and $99\pm3\rm\ \mu m$ for the 850 and $450\rm\ \mu m$ detectors, respectively.
\item The optimum surface impedance is $377\ \Omega/\square\,$. The absorption is not highly sensitive to the surface impedance.  In order to attain at least 99\% of the optimum efficiency the surface impedance must be between 309 and $460\ \Omega/\square\ $ for the $850\rm\ \mu m$ detectors and between 308 and $461\ \Omega/\square\,$ for the $450\rm\ \mu m$ detectors.
\item The monochromatic efficiency of the detectors at the optimum slab thickness is 94\%.
\item When we take into account the finite bandwidth of the filters, the efficiency of the detectors at the optimum slab thickness is 93\%\ at $850\rm\ \mu m$ and 88\%\  at $450\rm\ \mu m$. 
\item The detector efficiency is not very dependent on incident angle up to about $30\arcdeg$.  There is no significant loss of sensitivity over the range of incident angles in the field of view.
\item The flatness requirement for a decrease in efficiency of less than 5\%\ is that the slab thickness varies across a pixel by no more than $15\rm\ \mu m$.
\item The presence of a $1\rm\ \mu m$ oxide layer on the walls and slab has a negligible effect on efficiency.
\item The removal of the mechanical support grid improves the detector efficiency significantly.
\end{itemize}

We conclude from this modeling that the SCUBA-2 detectors will meet the requirement of an absorption efficiency of 75\%\ or greater.

\acknowledgments     
 
SCUBA-2 is a collaboration between the UK Astronomy Technology Centre (Edinburgh), the National Institute for Standards and Technology (Boulder), the Scottish Microelectronics Centre (Edinburgh), the University of Wales (Cardiff), the Joint Astronomy Centre (Hawaii), the University of Waterloo, the University of British Columbia (Vancouver), the University of Lethbridge, Saint Mary's University (Halifax), and Universit\' e de Montr\' eal.  The project is funded by the UK Particle Physics and Astronomy Research Council, the JCMT Development
Fund and the Canada Foundation for Innovation.  HFSS\texttrademark\ and Optimetrics\texttrademark\ are trademarks of Ansoft Corporation.  We would like to thank Kelvin Clarke at Ansoft for his help in troubleshooting the HFSS models.


\bibliography{SCUBA-2}   

\begin{thebibliography}{1}

\bibitem{SPIE04}
M.~D. Audley, W.~S. Holland, T.~Hodson, M.~J. MacIntosh, I.~Robson, K.~D.
  Irwin, G.~C. Hilton, W.~D. Duncan, A.~Walton, W.~Parkes, P.~A.~R. Ade,
  I.~Walker, M.~Fich, J.~Kycia, M.~Halpern, D.~A. Naylor, G.~Mitchell, and
  P.~Bastien, ``An update on the {SCUBA}-2 project.'' SPIE 5498, in press,
  2004.

\bibitem{Hadl47}
L.~N. Hadley and D.~M. Dennison, ``Reflection and {T}ransmission {I}nterference
  {F}ilters {P}art {I}. {T}heory,'' {\em J. Opt. Soc. Amer.} {\bf 37}(6),
  pp.~451--465, 1947.

\bibitem{Carli81}
B.~Carli and D.~Iorio-Fili, ``Absorption of composite bolometers,'' {\em J.
  Opt. Soc. Amer.} {\bf 71}(8), pp.~1020--1025, 1981.

\bibitem{Park04}
W.~Parkes, A.~M. Gundlach, C.~C. Dunare, J.~G. Terry, J.~T.~M. Stevenson, A.~J.
  Walton, and E.~Schulte, ``Realization of a large area microbolometer sensor
  array for submillimetre astronomy applications: {SCUBA}-2.'' SPIE 5498, in
  press, 2004.

\end{thebibliography}
\bibliographystyle{spiebib}   

\end{document}